\documentclass[sigchi, authordraft, screen, timestamp=false, review=false]{acmart}

\usepackage{booktabs} 

\usepackage{tabularx}
\usepackage{multirow}

\usepackage{caption}
\usepackage{subcaption}

\setcopyright{none}

\acmDOI{}

\begin{document}
\title{Learning and Anticipating Future Actions During Exploratory Data Analysis}

\author{Ran Wan}
\orcid{1234-5678-9012}
\affiliation{%
  \institution{Washington University in St. Louis}
}
\email{wanran@wustl.edu}

\author{Roman Garnett}
\affiliation{%
  \institution{Washington University in St. Louis}
}
\email{garnett@wustl.edu}

\author{Alvitta Ottley}
\affiliation{%
  \institution{Washington University in St. Louis}
}
\email{alvitta@wustl.edu}

%
%
%


\begin{abstract}
The goal of visual analytics is to create a symbiosis between human and computer by leveraging their unique strengths.
While this model has demonstrated immense success, we are yet to realize the full potential of such a human-computer partnership.
In a perfect collaborative mixed-initiative system, the computer must possess skills for learning and anticipating the users' needs.
Addressing this gap, we propose a framework for inferring focus areas from passive observations of the user's actions, thereby allowing accurate predictions of future events.
We evaluate this technique with a crime map and demonstrate that users' clicks appear in our prediction set 95\% -- 97\% of the time.
Further analysis shows that we can achieve high prediction accuracy typically after three clicks.
Altogether, we show that passive observations of interaction data can reveal valuable information that will allow the system to learn and anticipate future events, laying the foundation for next-generation tools. 

\end{abstract}

%
%
%


\begin{teaserfigure}
  \includegraphics[width=\textwidth]{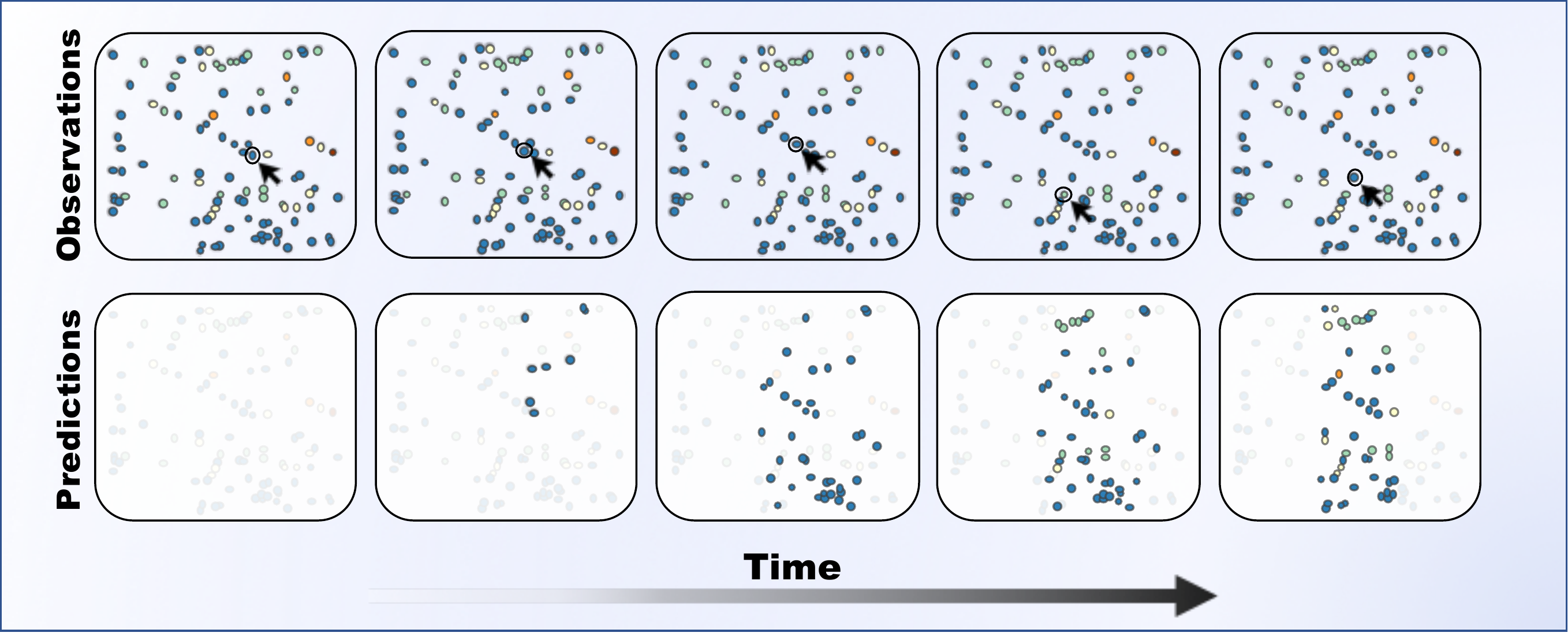}
  \caption{A simulation of the algorithm when applied to a scatterplot.
  	The simulated user begins by clicking on blue dots at the center of the visualization.
  	Within a few clicks, the algorithms' predictions for the user's evolving attention converge to circles of the representative color within the general area of the observed clicks. At $t=4$, the user selects a green dot in the same region, and subsequent predictions update to include circles of different colors and in a more tightly defined area.}
  \label{fig:teaser}
\end{teaserfigure}

\maketitle

\section{Introduction} 
\label{sec:introduction}
The overarching goal of \textbf{visual analytics} is to create a symbiosis between human and machine.
Visualization serves as a medium that allows users to collaborate with computers in ways that takes advantage of their distinct strengths~\cite{keim2008visual}.
Both Crouser and Chang~\cite{crouser2012affordance,crouser2013balancing} and Green et al.~\cite{green2008visual} describe an affordance-based partnership model that leverages the human's unique skills (e.g., reasoning and social awareness) with the machine's computational powers.
Typically, the human drives the analysis process by exploring the data to form hypotheses and develop insights.
Success in the analytic process hinges on the user's ability to perform meaningful interactions with the data and on the machine's ability to provide the right information at the right time~\cite{ellis2010mastering,keim2008visual}.

Although this model has shown remarkable success, for many analysts, the complaint \textit{``too much data -- not enough information''} is still all too common~\cite{newsom2013towards}.
In many ways, the tools fall short of their full potential.
A useful collaborative tool should possess the ability to learn about what the user is doing, what the user will be doing, what the user ought to be doing, and whether the current trajectory will solve the problem at hand.
Current visual analytics tools do not yet possess the ability to learn and anticipate actions, and therefore are unable to tailor their outputs.

These considerations have, in large part, driven the goals of many visual analytics researchers.
To understand what the user is doing, Pirolli and Card introduced the \textit{sensemaking loop,} which models an analyst's progression from information foraging through hypotheses generation and insight~\cite{pirolli1999information}.
Researchers have also created taxonomies for the types of tasks and interactions that are feasible for a given visualization~\cite{amar2005low,buja1996interactive,chuah1996semantics,dix1998starting,gotz2009characterizing,lee2006task,keim2002information,shneiderman1996eyes,wilkinson2006grammar,yi2007toward,zhou1998visual}.
Researchers have demonstrated automatic and manual techniques for tracking workflow~\cite{andrienko2018viewing,cowley2005glass,bavoil2005vistrails,dabek2017grammar,callahan2006vistrails,freire2006managing,heer2008graphical,javed2013explates}, analysis strategies~\cite{brown2014finding,ottley2015personality}, and personality~\cite{brown2014finding,ottley2015personality,steichen2013user,toker2013individual}.
To understand what the user ought to be doing, researchers introduced techniques for detecting cognitive biases for interaction data~\cite{cho2017anchoring,dimara2017attraction,wall2017warning}.
While these past efforts have demonstrated some success, predicting future events is still an open challenge, making it difficult to realize a human-computer team that genuinely operates in tandem.

The work in this paper builds on the prior results and aims to develop automatic techniques for learning and anticipated events during visual data exploration.
We propose a context-aware, data-driven prediction system that integrates advancements from artificial intelligence within a visualization tool to detect future interactions.
Specifically, we create a hidden Markov model that represents \textit{evolving attention} as a series of unobservable states giving rise to actions.
We can then automatically infer elements of interest from passive observations of the user's actions, thereby allowing accurate predictions of future interactions.

For a proof of concept, we conducted a controlled user study and collected click-stream data as participants explored a map visualization of reported crimes (see figure~\ref{fig:map}).
Our results show that the probabilistic model can achieve, depending on the type of task, between 95\% and 97\% accuracy at predicting future mouse clicks from observation of their click behavior.
Further analysis shows that we can achieve high prediction accuracy in a short period (typically after three clicks).
Altogether, we show that passive observations of interaction data can reveal valuable information about users' attention.

We posit that the work in this paper opens the door for many opportunities to improve analysts' experience and lay the foundation for next-generation visual analytics systems.
For instance, the machine can proactively perform tasks such as prefetching, calculation of summaries statistics, suggestion formation, bias or error identification, and target selection assistance for overcrowded interfaces.
We discuss how the proposed technique can help create next-generation visual analytics systems that can automatically learn users' focus to support the analysis process better.

We make the following contributions:
\begin{itemize}
	\item \textit{A design-agnostic approach to modeling interaction with visualization:} We provide a design-agnostic approach for automatically learning future event during data exploration and demonstrate, using a crime map, how to model users' interests and actions.

	\item \textit{Predicting future clicks from passive observations:} We demonstrate how to apply this model to a real-world visualization and dataset. Our proof-of-concept experiment validates that we can use this approach on real systems for real-time predictions. We demonstrate the participants' clicks appear in our prediction set on average 95\% of the time.

	\item \textit{Implications for designing mixed-initiative visualization tools:} We discuss techniques for supporting the user in real time and contribute to next-generation of visual analytics systems.

\end{itemize}

\section{Prior Work on Learning from Interaction Logs}
\label{sec:background}
Analyzing interactions to learn about the user or an interface design has been in important area of research across many fields.
For example, in machine-learning, researchers have used interaction data to model and predict users' browsing behaviors on websites and web search systems~\cite{eirinaki2003web, kolari2004web, kosala2000web, srivastava2000web}.
Some researchers have also used interaction data to explore how interface design can bias user behaviors~\cite{guan2007eye,joachims2017unbiased} and how to overcome these biases~\cite{joachims2017unbiased}.

In databases, Battle et al.~\cite{battle2016dynamic} analyzed interaction data to improve prefetching techniques. They showed that analyzing behavioral data resulted in a 430\% improvement in system latency.
In the HCI field, researchers showed that displaying interaction history of past users improves the problem-solving of future users~\cite{wexelblat1999footprints}.
Furthermore, Gajos et al.\ developed the SUPPLE system that can learn the type and degree of a user's disability by analyzing mouse interaction data~\cite{gajos2004supple,gajos2008decision,gajos2008improving}.
Fu et al. developed statistical and machine-learning models to predict behavior on crowdsourcing annotation and web search tasks~\cite{fu2017your}.
These are just a few of the many examples of related work, across a vast number of research communities.
However, most relevant to the the work in this paper is research in the area of \textit{analytic provenance}.

\subsection{Analytic Provenance}
It is a common belief that interaction logs contain crucial information about an analyst's reasoning process with a visualization~\cite{pike2009science}.
Through interaction with a visual interface, analysts explore data, form and revise hypotheses, and make judgments.
The term \emph{provenance} refers to the history of an object or idea, and \textit{analytic provenance} researchers aim to track and analyze the analytics process~\cite{freire2008provenance,ragan2016characterizing,nguyen2014survey,north2011analytic}.
At a high level, the goal is to automatically capture and encode interactions with a visual interface to infer analysts' goals and intentions.
Researchers and practitioners can then recall, replicate, recover actions, communicate, present, and perform meta-analyses on the analysis process~\cite{ragan2016characterizing}.

A standard approach to recovering the analytic process is to capture low-level user actions such as mouse and keyboard events.
For example, Cowley et al.\ developed \textit{Glassbox} with the goal of logging interactions to infer intent, knowledge, and work-flow automatically~\cite{cowley2005glass}.
Dou et al.~ demonstrated that is it possible to extract high-level information from interaction data~\cite{dou2009recovering}.
They conducted a user study and recorded interactions while financial analysts used a visual analytics system to detect wire fraud.
Through a manual analysis of the interaction data, they showed that is possible to recover analysts' strategies, methods, and findings.
More recent work by Feng et al.\ demonstrated metrics for quantifying the data exploration~\cite{feng2018patterns}.
Dabek and Caban introduced a grammar-based approach to modeling user interactions~\cite{dabek2017grammar}.
They used automatons to model users' behavior and demonstrated that their technique could capture user's analytic process.


A series of work focused on recording, annotating, and maintaining interaction history, and demonstrates the benefit of preserving a linear history for future use~\cite{bavoil2005vistrails,brodlie1993grasparc,callahan2006vistrails,gotz2006interactive,heer2008graphical,shrinivasan2008supporting}.
\textit{VisTrails}, for instance, automatically keeps track of the analyst's workflow and pipeline, making it possible for the user to resume, reuse, and share their explorations~\cite{bavoil2005vistrails,callahan2006vistrails,freire2006managing}.
Heer et al.~\cite{heer2008graphical} and Javed and Elmqvist~\cite{javed2013explates} created graphical history tools that would allow users to track, recall, and share their process.
Gotz et al.~developed tools for supporting the sensemaking process by augmenting existing data with user annotations~\cite{gotz2006interactive}.

\subsection{Analyzing Interaction to Infer User Attributes}
Researchers have also used interaction logs to infer user knowledge or intent.
Brown et al.\ used \textit{Dis-Function} to learn analysts' knowledge through direct manipulation of visual elements~\cite{brown2012dis}.
Users expressed their domain knowledge by grouping similar points. 
The system then used this information to update the underlying distance function for the data projection.
Prior work also demonstrates how interaction data can be used to steer computation and refine model parameters~\cite{endert2012semantic,endert2012semantic2,garg2008model,paurat2014interactive,saket2017visualization,sarvghad2018embedded,xiao2006enhancing}.
For example, Endert et al.~\cite{endert2012semantic,endert2012semantic2} designed \textit{ForceSPIRE,} which is a text data analysis tool that automatically updates the underlying layout model as users interact with documents.
Guo et al.\ analyzed interaction logs to understand how analysts achieve insights~\cite{guo2016case}.

Other researchers analyzed interaction data to infer individual characteristics.
For instance, recent work by Wall et al.\ introduced a framework for quantifying different types of biases and proposed a Markov chain technique for identifying biases in real time~\cite{wall2017warning}.  
Work by Brown et al.\ used machine-learning techniques to infer user attributes automatically~\cite{brown2014finding}.
They showed that off-the-shelf algorithms could successfully predict completion time and personality traits based on low-level mouse clicks and moves~\cite{brown2014finding}.
They also demonstrated the viability of making real-time inferences from passive observations.
Ottley et al.\ analyzed clickstream data to demonstrate a correlation between personality traits and search strategies with hierarchical visualizations~\cite{ottley2015personality}.
Lu et al.\ used eye-tracking data to select parameters for a visualization automatically~\cite{lu2010volume}.
Also utilizing eye gaze data, Steichen et al.~\cite{steichen2013user} and Toker et al.~\cite{toker2013individual} predicted cognitive traits such as visual working memory,  personality, and perceptual speed.

\section{General Modeling Framework}
\label{sec:model}
\begin{figure*}
	\centering
	\includegraphics[width=\textwidth]{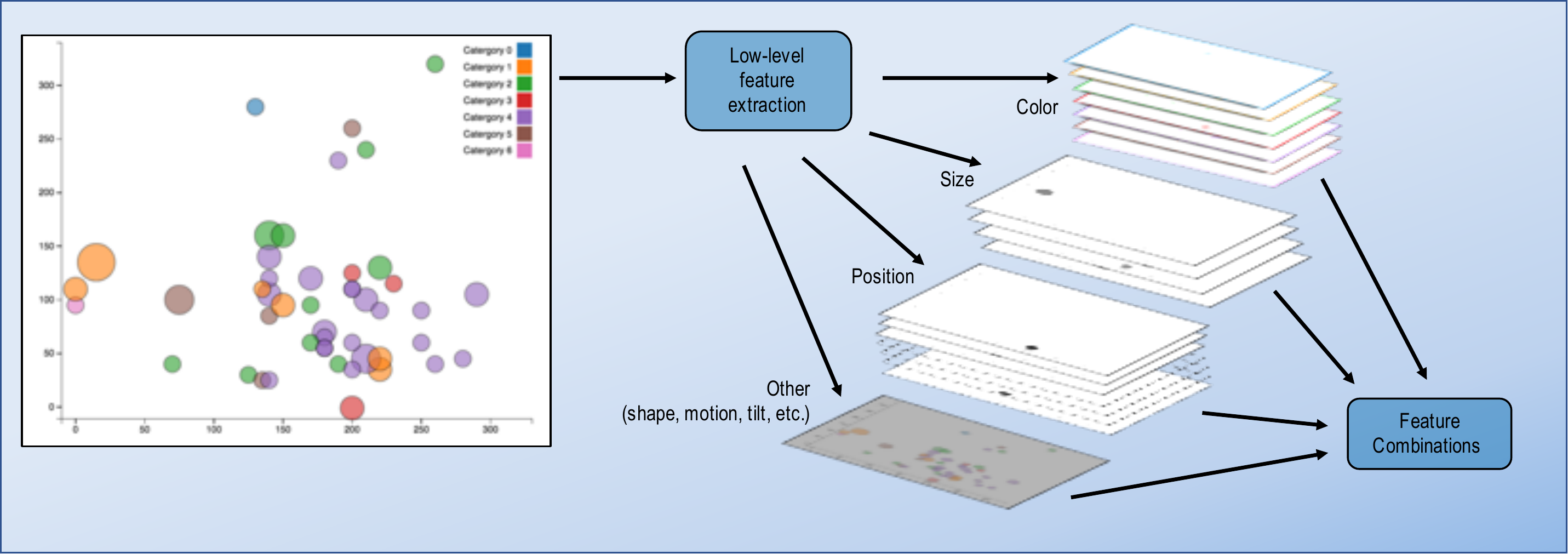}
	\caption{Extracting low-level features.}
	\label{fig:features}
\end{figure*}

The previous section recaps prior work aimed at learning from interaction data.
Much of the proposed approaches in the visualization community have primarily focused on analyzing behavior for tracking analytic provenance.
In this work, we propose and demonstrate, for the first time to the best our knowledge, a model for predicting mouse interactions before they occur.

The goal is to create a computational model that is task- and design-agnostic.
We use a bottom-up approach that utilizes low-level visual features that we extract from the visualization design (e.g., color, shape, and position).
We make passive observations of low-level interaction and consider the properties associated with each visual element the user interacts with.
It is important to note that low-level features do not incorporate top-down signals that may be derived from the task at hand.
However, the extraction process (detailed in figure~\ref{fig:features}) is robust and has proved successful for modeling visual attention in images~\cite{itti1998model,itti2000saliency,itti2001computational,koch1987shifts}.

\begin{figure}[b]
	\centering
	\includegraphics[width=.5\textwidth]{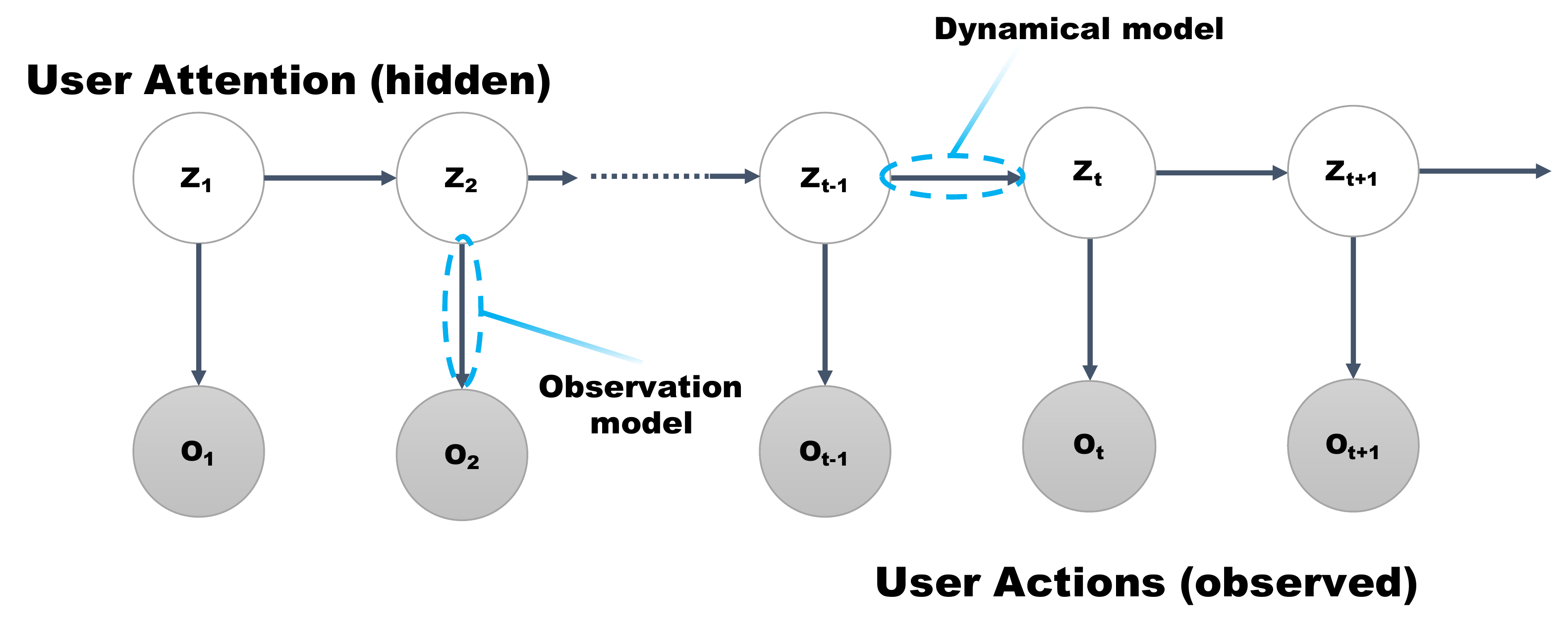}
	\caption{A hidden Markov model approach to modeling attention and actions with a visualization system. We represent evolving attention as a sequence of latent variables in the hidden state space. Observable states are the user's actions. The conditional distribution of each observation depends on the state of the corresponding latent variable.}
	\label{fig:hmm}
\end{figure}

We construct a \emph{hidden Markov model,} presuming the user's attention evolves under a Markov process (that is, the attention at a particular time only depends on their attention at the previous time step), and interaction events are generated conditionally independently given this sequence of attention shifts.
Figure~\ref{fig:hmm} shows an overview of the hidden Markov model used.
We represent selective attention as a sequence of latent variables.
The conditional distribution of each observation depends on the state of the corresponding latent variable.
To specify this model, we need to define the following:
\begin{itemize}
	\item \textbf{Unobservable states:} A space of the possible "interests" driven by the salient visual features.
	\item \textbf{Observable states:} A space of possible interactions.
	\item \textbf{Dynamical model:} A model of the evolution of the user's attention over time.
	\item \textbf{Observation model:} A model of how attention gives rise to observed actions.
\end{itemize}

\subsection{Defining Unobservable and Observable States}

First, we define a discrete time index $t$ associated with interactions with a visualization.
At the start of exploring the dataset, we define $t=0$.
This index will then increment every time a participant interacts with a visual element. Our model will presume that there is a \emph{hidden, unobserved} state $z_t$ representing the attention of the user at time $t$.
We will assume that we can map the sequence of observed interactions $\{o_t\}$ to this hidden sequence of focus areas.
The task we consider here is how to \emph{infer} the hidden attention/focus of the user by observing their sequence of interactions.

In order to create a model of user interaction, we must first understand the mechanisms that drives the user to interact with a particular visual element.
Our model assumes no expertise or prior knowledge from the user.
We also assume that innate biological models of selective attention drive interactions.
At a high level, we build on Koch and Ullman's model of visual attention~\cite{koch1987shifts} and learn a \textit{saliency map} for a given time step.

\subsubsection{\textbf{Unobservable States}}
We therefore begin by segmenting the visualization based on the low-level visual features.
We define $\mathcal{M}$ as the mark space that specifies the types of visual marks and channels used in the visualization.
Visual marks are geometric elements, and there are four primitive types: points, lines, areas, and volumes~\cite{bertin1983semiology}.
Visual channels describe the graphical properties of visual marks such as position, size, color, luminance, shape, texture, and orientation~\cite{bertin1983semiology}.
Together with Card et al.'s data-mapping principles~\cite{card1999readings} these design guidelines can be used to describe any existing visual representation~\cite{card2009information}.
We create $\mathcal{M} = \{f_1, ..., f_N\} $ by decomposing the visualization into its primitive visual marks and channels, as detailed in figure~\ref{fig:features}.

A crucial component of the probabilistic model is the specification of a hidden state space, which will represent the attention of the user at a given time.
In general, we propose that designers can tailor this space for a given scenario.
In many scenarios, we may reasonably assume the users' attention at a given time to be related to some weighted subset of visualization marks, for example, visual marks of a particular size, color, shape, or in a specific location.
In such a case we may define the latent attention at time $t$, as $z_{t} = \{f_{1_{t}}, ..., f_{N_{t}}, \pi_{t}\}$ where $\pi$ represents the feature weights,
and $\{f_{1_{t}}, \dotsc f_{N_{t}}\}$ represent feature values describing the user's focus at time $t$.
We provide more details for the feature weights below.


\subsubsection{\textbf{Observable States}}
In contrast to the hidden attention space, the space of observed actions is typically easy to define.
We may define $o_t$ to be an observation of the user at time $t$, where this observation will be an interaction event with a visual element (e.g., mouse clicks, mouse moves, eye gaze, etc.).
We will represent each observation $o_{t} = \{f'_1, ..., f'_N\} $ as the set feature values that describes the visual element.

\begin{table}[t]
	\footnotesize
	\centering
	\caption{Mathematical symbols.}
	\label{tab:symbols}
	\begin{tabularx}{\columnwidth}{cX}
		
		\textbf{Symbols} & \textbf{Description}   \\
		\hline
		$t$                                   & the time an event occurs.\newline \\
		$\mathcal{M} = \{f_{1}, ..., f_{N}\} $      &  Mark Space: The set of N visual features extracted from the visualization (e.g., position, size, and color). \newline \\
		$o_{t} = \{f'_1, ..., f'_N\} $ & observation interaction at time $t$ (e.g., click, gaze, and hover). We consider set of values for the N features.\newline \\
		$\pi = [ \pi(f_{1}), ..., \pi(f_{N})]$              & bias vector for all features $f \in  \mathcal{M}$. \newline \\
		$z_{t} = \{f_{1_{t}}, ..., f_{N_{t}}, \pi_{t}\} $ & latent attention at time $t$.\newline \\
		\hline
	\end{tabularx}
\end{table}

\subsection{Dynamical Model}

The full specification of a hidden Markov model requires defining a probabilistic model of the dynamics of the hidden state space, that is how the user's latent attention shifts from one time-step to the next.
We define $z_t$ to be the latent attention of the user at time $t$.

\subsubsection{\textbf{Single Task}}
We model shifts of attention by defining a probability distribution $p(z_{t + 1} \mid z_t)$ describing the evolution of attention.
We propose that this model should be reasonably easy to define in most visualization settings.
In general, it is unlikely that the user's focus will change rapidly from one interaction event to the next.
Therefore we can often choose this dynamics model to represent a simple random diffusion in the latent space:
\[
  z_{t + 1} = z_t + \varepsilon,
\]
where $\varepsilon$ is some appropriate noise distribution (e.g., zero-mean Gaussian noise for real-valued features or a discrete distribution favoring $z_{t + 1} = z_t$ for discrete features, see also below).
This model assumes that focus of attention is likely to remain constant from time $t$ to $t+1$, with some slow decay as the user continues to interact with the system.
This is consistent with psychological research that suggest that selective attention does not change drastically over time~\cite{koch1987shifts}.

\subsubsection{\textbf{Multiple Tasks}}
If a visualization setting may comprise a sequence of separate tasks, we may also construct dynamical models that loosely encode that user's attention may change in one of two ways: either the current task has not yet completed, in which case we may assume a simple drift model as described above.
Otherwise, if the task has completed, we might model the attention at the next time step as being drawn from some broad distribution over the space of possible focus points.
In such a construction our dynamical model would be a mixture distribution with two components corresponding to the continuation of a task or beginning a new task.
Such an approach has been used to model user intent in online games from observed low-level behavior~\cite{garnett_et_al_cig_2014}.

\subsubsection{\textbf{Bias}}
Koch and Ullman hypothesized that it is useful to consider bias when modeling attention shifts~\cite{koch1987shifts}.
Similarly, recent work by Wall et al.~\cite{wall2017warning} proposed a framework modeling different types cognitive biases during visual data exploration.
Motivated by the prior work, we adopt a bias vector $\pi = [ \pi(f_1), ..., \pi(f_N)]$ to capture the relative importance of the various components of the mark space where  $\pi(f) \in [0, 1]$.

\subsubsection{\textbf{Evolution of Attention}}
\label{sec:evolution}
For the dynamical model of the hidden state $p(z_t \mid z_{t-1})$, we assume that the attention at time $t + 1$ is typically similar to the attention at the previous time step $t$; that is, that attention does not change rapidly over time. We further assume that the each component of the attention vector evolves independently:
\[
p(z_{t+1} \mid z_t)
=
p(f_{1_{t+1}} \mid f_{1_{t}})
...
p(f_{N_{t+1}} \mid f_{N_{t}})
p(\pi_{t+1} \mid \pi_t).
\]

\paragraph{Continuous Features}
For an arbitrary continuous feature $f$ such as position, we may model of evolution the features using additive zero-mean Gaussian noise:
\[
  p(f_{t + 1} \mid f_t, \sigma_f^2) = \mathcal{N}(f_{t + 1}; f_t, \sigma_f^2),
\]
where the parameter $\sigma_f^2$ is the variance of the drift. For strictly positive values such as size or intensity, we could use a similar diffusion on the logarithm of the value instead, or we could simply project onto the feasible domain.

\paragraph{Categorical Features}
One possibility for modeling the evolution of an arbitrary discrete parameter $f$ such as color or shape is a simple ``biased coin flip'' model favoring no change:
\[
  p(f_{t + 1} \mid f_t, \rho)
  \rho \delta(f_t) + (1 - \rho) \mathcal{U}_{\setminus f_t},
\]
where $\rho$ is a parameter modeling the fickleness of the user, $\delta(f_t)$ is the Kronecker delta distribution with support $f_t$, and $\mathcal{U}_{\setminus f_t}$ is the uniform distribution over the values not equal to $f_t$. This distribution effectively says the user's attention does not change with probability $\rho$; otherwise, it changes to a different value with equal probability.

\paragraph{Ordinal Features}
We suggest treating ordinal feature as either categorical or continuous and using one of the above.

\subsubsection{Bias}
We also suggest that the relative importance of the various components of the mark space should remain relatively stable over time, and can adopt a diffusion for the bias parameter $\pi$ as well:
 \[
   p(\pi_{t + 1} \mid \pi_t) = \mathcal{N}(\pi_{t + 1}; \pi_t, \sigma_\pi^2).
\]
Note that we must account for boundary effects and normalization effects when defining the dynamical model; in practice, we may simply project out-of-range values onto their feasible domains.

\subsection{Observation Model}

We must also specify an observation model $p(o_t \mid z_t)$, which defines how latent user attention generates interactions.
We must take care to define such an observation model appropriately for a given scenario, and we will demonstrate how we might construct an explicit example in our user study below.
In a visualization setting, defining a reasonable choice for such a model is relatively straightforward.
If a user's attention is represented by some values in the same space as the visual elements in the visualization, we may often construct an observation model that loosely specifies that ``users interaction with related to their hidden focus space.''
We will show an explicit construction of such a model in Section~\ref{sec:casestudy}.


\subsection{Predicting Movement}
\label{sec:particle-filter}
Our goal at each time stamp is to predict the user's possible next interactions given the set of the user's previously observed events.
To approach this goal, we will use our hidden Markov model to infer the attention of the user at time $t$, $z_t$, given the interactions up to time $t$, $E_t = \{e_i\}_{i=1}^t$.
Unfortunately, this inference is usually not possible in closed form, but we can use a particle filter.
Particle filtering is a well-established technique for inferring the hidden states of dynamical systems such as ours~\cite{doucet2000sequential,gordon1993novel}.

We represent our belief about the latent state $z_t$ given the previous events $O_t$ with a set of $m$ \emph{particles} $\{z_t^{(i)}\}_{i=1}^m$, each particle a point in the attention space.
These particles represent samples from the posterior distribution $p(z_t \mid O_t)$.
Suppose for induction that we have a set of such particles.
Particle filtering proceeds by repeating the following steps:
\begin{itemize}
    \item We push the particles through the dynamical model $p(z_{t + 1} \mid z_{t})$ by sampling a new value for each particle:
    \[
    z_{t+1}^{(i)} \sim p(z_{t+1} \mid z_t = z_t^{(i)}).
    \]
    \item
    We observe the next interaction event $e_{t+1}$ and weight the particles
    according to the agreement with the observation by evaluating the observation model:
    \[
    w^{(i)} = p(o_{t+1} \mid z_{t+1} = z_{t+1}^{(i)}).
    \]
    \item
    We sample a new set of $m$ particles by sampling with replacement
    from the set of existing particles with probability equal to the
    weights $\{w^{(i)}\}$.
\end{itemize}
This set of resampled particles will represent a sample from the distribution
$p(z_{t+1} \mid O_{t+1})$, and we may proceed inductively.

For each timestamp, we can get $p(z_t \mid O_t)$, which is the particle given all previous interaction events. However, particles can be at any location on the visualization. Our goal is to find possible visual element users are going to interact with at the next time stamp.

To do this, we need one extra step. We treat every mark on the visualization as a potential candidate for the next interaction.
We sum the weight every particle contributes to that candidate using the observational model. A subset of marks with highest weights, $\alpha$, will be considered as predictions. Notice that the size of $\alpha$ at this point is arbitrary.


\section{Example Application }
\label{sec:casestudy}
\begin{figure}[h]
	\centering
	\includegraphics[width=\columnwidth]{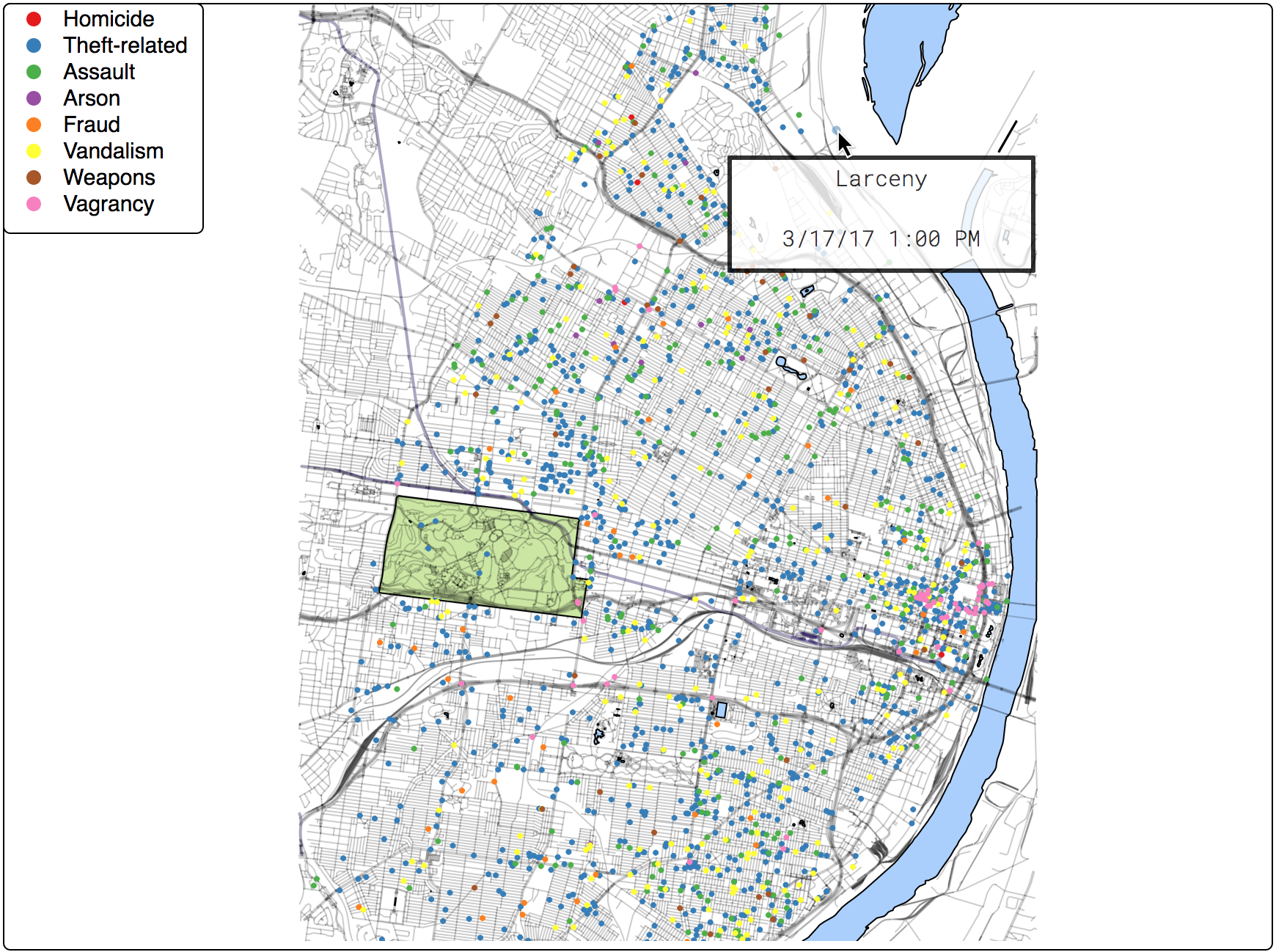}
	\caption{The interface used in our experiment. Participants used their mouse to pan and zoom the map. A tooltip displayed information about the crimes on click. }
	\label{fig:map}
\end{figure}

We now apply this model to a visualization interface (see figure~\ref{fig:map}).
We chose a map for our study because of its broad application and use.
Below we demonstrate how to define the hidden state space and discuss choices for the dynamical and observation models.
In this example, we assume that users interact with visual marks by clicking on them.

\hspace{-20mm}

\subsection{Defining Unobservable and Observable States}

We define $o_t$ to be the click event at time $t$, which we will represent as a three-dimensional vector $o_t = (x'_t, y'_t, k'_t)$, where $(x'_t, y'_t)$ is the x-coordinate and y-coordinate of the click and $k'_t$ is the color of the circle clicked, represented by a discrete integer-valued index ranging over the eight possible values $\{1, ..., 8\}$. Note that we use prime symbols to indicate quantities associated with a click event.

Next, we will define a hidden state space modeling the attention of the user. Each point in this hidden space is a vector specifying (1): a location $(x, y)$ of interest, (2): a mark color $k$ of interest, and (3): a bias parameter indicating the relative importance of location and mark color. For this example, we represent the bias parameter as a number $\pi \in [0, 1]$, with 1 indicating a complete focus on location and 0 indicating a complete focus on mark color. A point in this latent attention space is thus a four-dimensional vector $z = (x, y, k, \pi)$.

Our model assumes that at every discrete time step $t$ in the interaction process (each time the user makes a click), the user has an underlying attention $z_t$ corresponding to a vector in the attention space defined above. We seek to infer the attention of the user through observing the sequence of click events $\{o_t\}$. We will approach this inference problem via creating a hidden Markov model and performing inference with particle filtering.

Our model is fully specified by a dynamical model $p(z_t \mid z_{t-1})$ describing how the hidden state evolves and an observation model $p(o_t \mid z_t)$ describing how a hidden attention vector generates click events. We define each of these below.

%
%
%
%
%
%
%

\subsection{Dynamical Model}
\label{sec:dynamical-model}
Here, we adopt a simple stationary diffusion model. As detailed in Section~\ref{sec:evolution}, we assume that the four components of the attention vector evolve independently:
\[
p(z_{t+1} \mid z_t)
=
p(x_{t+1} \mid x_t)
p(y_{t+1} \mid y_t)
p(k_{t+1} \mid k_t)
p(\pi_{t+1} \mid \pi_t).
\]

We model the evolution of the continuous location and location--color bias parameters with a simple Gaussian drift:
\begin{align*}
p(x_{t+1} \mid x_t, \sigma_x)
&=
\mathcal{N}(x_{t+1}; x_t, \sigma_x^2);
\\
p(y_{t+1} \mid y_t, \sigma_y)
&=
\mathcal{N}(y_{t+1}; y_t, \sigma_y^2);
\\
p(\pi_{t+1} \mid \pi_t, \sigma_\pi)
&=
\mathcal{N}(\pi_{t+1}; \pi_t, \sigma_\pi^2).
\end{align*}
The expected value of these parameters is equal to the previous value, with zero-mean Gaussian diffusion with parameter-dependent variance added. We will select these parameters $\sigma_{x}$, $\sigma_{y}$, and $\sigma_{\pi}$. Notice also that these three parameters are all also bounded values: the locations $x$ and $y$ indicates a position on the map and must lie in its domain, and the bias parameter $\pi$ must lie in the interval $[0, 1]$. Therefore, we need to deal with cases when the diffused value steps outside the boundary. Here we simply adopted a rule that whenever a diffused value steps outside the boundary for a variable, we move it onto the boundary in the direction of diffusion. For example, if $\pi_{t+1}$ diffuses to value greater than 1, we will set it to 1; likewise if the diffused location $(x_{t+1}, y_{t+1})$ lies beyond the width and height of the map, we will project onto the nearest point on the canvas boundary.

Lastly, because mark color is a categorical value, we cannot directly apply normal diffusion to it. Here we used a discrete analog of that diffusion following our suggestion in Section~\ref{sec:evolution}. We define a transition probability $\rho$ and assume that with probability $\rho$ the latent mark color of interest does not change. Otherwise, a new mark color of interest is chosen from all possible values with equal probability:
\[
p(k_{t+1} \mid k_t, \rho) =
\rho\delta(k_{t}) + (1-\rho)\mathcal{U}_{\setminus k_t},
\]
where $\delta$ is a Kronecker delta distribution and $\mathcal{U}(K)$ is a uniform distribution over the mark colors except $k_t$. Again this choice models our assumption that attention typically changes slowly over time.

\subsection{Observation Model}
\label{sec:observation-model}
We must also specify an observational model $p(o_t \mid z_t)$ modeling the probability of a click event $e_t = (x'_t, y'_t, k'_t)$ given the attention $z_t = (x_t, y_t, k_t, \pi_t)$ at time $t$. A brief summary of this observational model is that we flip a coin with heads probability equal to the location--color bias parameter $\pi_t$. If the coin lands heads, we assume the user is focusing on location and will probably click somewhere near the location in $(x_t, y_t)$. If not, we assume the user is focusing on mark color and will click on a mark of the color $k_t$. Specifically, we define:
\begin{multline*}
p(e_t \mid z_t, \sigma_x, \sigma_y)
=
\\
\pi
\mathcal{N}(x'_t; x_t, \sigma_x^2)
\mathcal{N}(y'_t; y_t, \sigma_y^2)
+
(1 - \pi)
\mathcal{U}(k'_t; k_t),
\end{multline*}
where $\mathcal{U}(k'_t; k_t)$ denotes a uniform distribution over the available marks of color $k_t$. This above model therefore assumes that that if the user is interested in position (with probability $\pi_t$), she will click on a position on the map with probability proportional to a Gaussian distribution centered on $(x_t, y_t)$ with diagonal covariance $[\sigma_x^2, 0; 0, \sigma_y^2]$. Again, we will specify these parameters.


\subsubsection{Predicting Movements}
To prediction movements, we can apply particle filter as described in Section~\ref{sec:particle-filter}.
Figure~\ref{fig:teaser} shows an simulation of the algorithm when applied to a simple scatter plot.
The simulated user begins by clicking on blue dots at the center of the projection, and within a few clicks, interest predictions converge to circles of the representative color with similar locations. At $t=4$, the user selects a different color circle in the same region, and subsequent predictions update to include circles of different colors in a more tightly defined area.

\section{Evaluation}
\label{sec:experiment}
To test our approach, we designed a user study to track and analyze interactions.
The dataset presented on the map were reported crimes in the city of St.\ Louis for March 2017 and that we gathered from the St.\ Louis Metropolitan Police Department's database~\cite{stlmpd}.
The dataset contained 20 features and 1951 instances of reported crime with eight different categories: Homicide, Theft-Related, Assault, Arson, Fraud, Vandalism, Weapons, and Vagrancy.

To visualize the crime instances, we used a single visual mark (we represented each crime as a circle on the map). The visual channels used were position and color which denoted the location and type of crime respectively. To separate intentional from unintentional interaction,  users interacted with the map by clicking on crime instances which triggered a tooltip displaying information about the type of crime and when it occurred.

\subsection{Participants}
We recruited 30 participants via Amazon's Mechanical Turk.
Participants were 18 years or older and were from the United States.
Each participant had a HIT approval rate greater than 90\% with at least 50 approved HITs.
We paid a base rate of \$1.00, an additional \$0.50 for every correct answer plus \$1.00 for each of the two optional post-surveys they completed.
The maximum reward was \$6.00.

There were 17 women and 13 men in our subject pool with ages ranging from 21 to 56 years ($\mu = 33.5$ and $\sigma = 10$).
Sixty percent of the participants self-reported to have at least a college education.

\subsection{Task}
In the main portion of the study, participants interacted with the crime map through panning, zooming and clicking to complete six search tasks and their associated question.
We divided these questions into three different task conditions. The three question types were meant to represent simple lookup tasks for which the participant had to consult the visualization:

\begin{itemize}
    \item \textbf{Geo-Based}: Different types of crime that are constrained to a specific geographical region.
    \item \textbf{Type-Based}: Same types of crime across the entire map.
    \item \textbf{Mixed}: Same types of crime \textbf{and} constrained to a specific geographic region.
\end{itemize}

The questions were simplified versions of real-world tasks that represented a potential interest. For instance, a person who in interested in buying a house may visit a crime map to learn about the types of crimes that frequently occur in the neighborhood (Geo-Based). A fire marshal may be interested in trends across reported cases of Arsons (Type-Based), or an investor may want to learn about theft crimes that tend to occur near a potential business site (Mixed).

The Geo-Based questions asked the participants to count the number of crimes within a specified geographical location that had a specific property.
For example, \textit{``Count the number of crimes that occurred during AM in the red-shaded region.''}
Participants clicked on every crime instance (a total of 43 dots) in the specified region.
They then chose their response from a series of multiple choice options.


Unlike the Geo-Based questions, the Type-Based tasks were not bounded to a specific region. These questions required participants to explore the entire map and search for a specified category of crime. For instance, \textit{``How many cases of Arson occurred during PM?''} To answer the question correctly, the participant would click on each instance of Arson (a total of 14 violet dots) to count the number of cases that occurred during PM.

For Mixed tasks, participants interacted with points of the same category of crime in a specified area. For example, \textit{``There are four types of Theft Related Crimes: Larceny, Burglary, Robbery, and Motor Vehicle Theft. Count the number of cases of Robbery in the red-shaded region.''}
Participants clicked on blue dots in the red-shaded area to reveal the tooltip (a total of 85 dots) and recorded the instances of Robbery.

While we used the same dataset throughout the experiment, each task focused on a different area of the map and a different type of crime.
To correctly answer the questions, the design of the task required the participant to click on every valid point in the dataset.  This was done to ensure a reasonably rich and large interaction dataset.

\subsection{Procedure}
After selecting the task on Mechanical Turk, participants consented per [redacted for anonymity] IRB protocol. They read the instructions for the study, then the main portion of the study began with a short video demonstrating the features of the interface. Specifically, we showed instructions for panning and zooming, and how to activate the tooltip.  The participant then completed the six search tasks and entered their answers for each by selecting the appropriate multiple choice response. The order of the six tasks was counterbalanced to prevent ordering effects. Once the tasks were done, they completed a short demographic questionnaire.

\subsection{Data Collection and Cleaning}

During the experiment, we recorded every mouse click event. We tracked the data point, its coordinates and a timestamp for the mouse event.
Each participant completed 6 tasks (two per task type), resulting in 180 trials.
To ensure the best quality data for our analysis, we filtered participants with incorrect answers.
We further removed tasks with less than five mouse click events.
After cleaning, 78 trials remained (28, 23, and 27 trials for Geo-Based, Type-Based and Mixed tasks respectively).

\begin{figure}[b]
	\centering
	\includegraphics[width=\columnwidth]{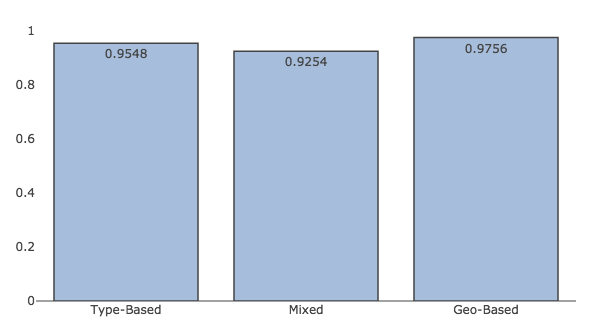}
	\caption{The average prediction accuracy across the three type of tasks. For $\alpha = 100$ our algorithm successfully predicted the users' next click, on average, 95\% of the times. }
	\label{fig:100accuracy}
\end{figure}

\subsubsection{Predicting Movement}
To predict movements, we applied particle filter as described in Section~\ref{sec:particle-filter}.
Although the choice for $\alpha$ (the size of the prediction set) can be adapted for the application, our goal was to have a prediction set that is small, relative to the size of the dataset.
We set $\alpha = 100$ which represents 5\% of the dataset used in the study.
This means that for a given timestep $t$, the algorithm chooses 100 points with the highest likelihood of being clicked at $t+1$.



\subsubsection{Parameters}
We used 1000 particles. The parameters were set as $\sigma_x = \sigma_y = 0.1, \sigma_\pi = 0.45$. The location scale parameters were again a fraction of the width and height of the map. The probability of maintaining the same type of crime as the users' attention $\rho$ was defined to be $0.96$.

\subsection{Results}
\subsubsection{Prediction Accuracy}
After gathering the data, we analyzed our model's ability to observe mouse clicks and predict interactions before they occur.
To allow time for the algorithm to learn users' attention, we begin our predictions at $t = 3$.
If the click at $t+1$ falls within our prediction set, we consider this a success. 
For each type of task (Geo-Based, Type-Based, and Mixed) we measured the overall predictive accuracy across all available clicks for all users:

\[\frac{\sum  successfulPredictions}{\sum predictions}\]

Figure~\ref{fig:100accuracy} shows the model's accuracy for each of the three tasks.
For $\alpha = 100$ (5\% of the dataset), our technique attained an average of 95\% at predicting the users' next clicks across all three task type ($\mu = .9548$, $\sigma = .1245$ for Type-Based, $\mu = .9254$, $\sigma = .0485$ for Mixed, and $\mu = .9756$, $\sigma = .0719$ for Geo-Based tasks).
In other words, with high accuracy, we can predict that the next click will be within a small set of data points, relative to the dataset.

\begin{figure}[h!]
\centering
\includegraphics[width=\columnwidth]{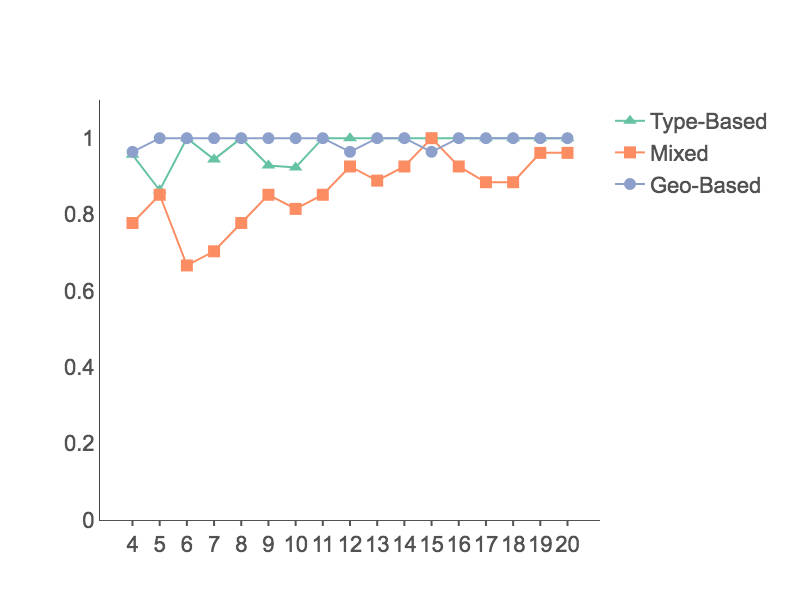}
\caption{The average accuracy over time for the three types of tasks in the study. After learning from 3 click interactions, the algorithm immediately achieves high prediction accuracy. We found that prediction accuracies remain fairly constant over time. }
\label{fig:accuracy-time}
\end{figure}

\subsubsection{Accuracy Over Time}
Our second analysis sought to evaluate our algorithm's performance as a function of the number of clicks observed.
For each type task, we measure the prediction accuracy (set size = 100) for the first 20 clicks observed.
Consistent with our previous analysis, we begin our predictions at $t = 3$.
Figure~\ref{fig:accuracy-time} summarizes our findings.
Our analysis reveals that the technique promptly achieves high prediction accuracies and performance remains fairly constant with more observations.

\section{Discussion \& Future Work}
\label{sec:discussion}

The hidden Markov model is a general framework that is widely used for modeling sequence data in areas such as natural language processing~\cite{manning1999foundations}, speech recognition~\cite{jelinek1997statistical,rabiner1993fundamentals}, and biological sequencing~\cite{durbin1998biological,sonnhammer1998hidden}.
However, we demonstrate its utility for modeling interest from interaction with a visualization system.
There are many possible variations for the model, the implementation, and parameters settings.
Examples include choices for the diffusion parameters, number of particles for the particle filter, and prediction set sizes.
A designer may tune these parameters or customize them based on the visualization or task.
We see this as a strength of the approach which can seed many opportunities for future work.

Although, the evaluation uses a single interface, we posit that the approach in this paper is generalizable under transparent assumptions.
We leverage data mapping principles and the notion that we can represent a visualization as a set of primitive visual marks and channels.
Designers can apply the approach to any visualization that can be specified in this manner.
The model assumes that the visual marks are perceptually differentiable, and relies heavily on good design practices.
To specify a user's evolving attention, we must first carefully define the mark space, $\mathcal{M}$.
One way to improve this process is to automatically extract the visual marks and channels from the visualization's code. However, this is beyond the scope of the paper.

Modeling attention can be a rich signal for inferring goals, intention and interest~\cite{horvitz1999principles,horvitz2003models}, and information about users' current and future attention can be useful for allocating computational resources~\cite{horvitz2003models} or for supporting data exploration. 
For example, the system can perform pre-computation or pre-fetching based on its predictions. 
For large datasets that may have overlapping points, a straightforward approach can be to redraw the points in the prediction set.
Doing so can make it easier for users to interact with points that match their interests but may have initially been occluded by other visual marks.
For more passive adaptations, designers can use the approach in this paper to inform techniques for target assistance~\cite{bateman2011target}.
The \textit{bubble cursor} technique, for example, does not change the visual appearance of the interface but increases the click radius for the given target, thereby making them more accessible~\cite{grossman2005bubble}.
Another possibility is \textit{target gravity,} which attracts the mouse to the target~\cite{bateman2011target}.
Future work can explore how to utilized to support the user during data exploration and analysis tasks. 

The general idea of mixed initiative systems~\cite{allen1999mixed,horvitz1999principles,horvitz1999uncertainty,horvitz2007reflections} or tailoring an interface based on users' skills or needs has existed for many years in HCI~\cite{gajos2004supple}. 
Researchers have explored the tradeoff between providing support and minimizing disruptions~\cite{afergan2013using,peck2014designing,solovey2011sensing,treacy2015designing}.
The work in the paper aligns well with this broader research agenda. 
We believe that the proposed approach is a significant step toward creating tools that can automatically learn and anticipate future actions, and opens possibilities for future work.

\subsection{Future Work}

One possible path for future work is to investigate the model's performance for more complex tasks.
In our experiment, we controlled the tasks by instructing participants to either search for a specific reported crime or identify a pattern in the dataset.
While these tasks were designed based on realistic scenarios, they assume that the user has a specific and unchanging goal when they interact with the visualization.
As a result, the search patterns we observed may not generalize to open-ended scenarios, or when the user's interest change while interacting with the data.
It is also possible that there are some scenarios where the user's attention cannot be represented at as subspace of the visualization marks (e.g., attending to negative space).
Future work can evaluate the approach with open-ended tasks.


The combination of visual marks and channels is an essential factor when defining the hidden state space for our probabilistic model. The map used in our experiment was simplistic compared to other real-world visual analytics systems.
Future work can test the model using different combinations of visual variables and channels on a single map, or an entirely different type of visualization.
It is also common for designers to aggregate the data based on the zoom level of the interface.
It is essential to validate the technique by changing and increasing the size of the dataset, which can result in the drastic changes in the appearance and number of visual marks.


\section{Conclusion}
\label{sec:conclusion}
In this paper, we have proposed a generalizable and design-agnostic approach to modeling users' evolving attention and actions with a visualization system.
We used a hidden Markov model and represented attention using the primitive visual marks and channels of the visualization design.
We demonstrated with a simple map how to apply this approach to a given visualization design.

To evaluate this technique, we conducted a user study and captured interaction data as participants explored a map showing a real-world crime dataset. 
The results of the study demonstrate that the approach is highly successful at modeling interaction and predicting users' next clicks. 
We observed an overall accuracy of 95\% at guessing actions before they occur.
These results are exciting and contribute to our overall goal of creating intelligent systems that learn about the user, her analysis process, and her task as she uses the system.
We believe that the work in this paper is a significant step toward this goal and can act as a catalyst for future work aimed at developing visual analytic systems that can better support users.

\begin{acks}

  The authors thank you for your valuable comments and helpful suggestions. 

\end{acks}

\bibliographystyle{ACM-Reference-Format}
\bibliography{modeling}

\end{document}